\documentclass{aa}  
\usepackage{graphicx}
\usepackage{natbib}
\bibpunct{(}{)}{;}{a}{}{,} %
\usepackage{txfonts}
\begin{document}
   \title{ADONIS high contrast infrared imaging of Sirius-B.}

   \titlerunning{Infrared imaging of Sirius-B}

   \author{J.M. Bonnet-Bidaud\inst{1,2}
   \and E. Pantin\inst{1,2}
   }
       
   \offprints{J.M. Bonnet-Bidaud}

   \institute{CEA, Irfu, Service d'Astrophysique, Centre de Saclay, 91191 Gif-sur-Yvette, France. 
    (\email{bonnetbidaud@cea.fr)}       
\and
Laboratoire AIM, CEA-CNRS-Universit\'e Denis Diderot, Paris, France}  
   \date{Received ; accepted }

 
  \abstract
   {Sirius is the brightest star in the sky and a strong source of diffuse light for modern telescopes so that the immediate surroundings of the star are still poorly known.}
   {We study the close surroundings of the star (2 to 25\arcsec) by means of adaptive optics and coronographic device in the near-infrared, using the ESO/ADONIS system. }
   {The resulting high contrast images in the JHKs bands have a resolution of $\sim$ 0.2\arcsec\, and limiting apparent magnitude ranging from m$_K$=9.5 at 3\arcsec\, 
from Sirius-A to m$_K$=13.1 at 10\arcsec. These are the first and deepest images of the Sirius system in this infrared range.}
   {From these observations, accurate infrared photometry of the Sirius-B white dwarf companion is obtained.  The JH magnitudes of Sirius-B are found to agree with expectations for a DA white dwarf of temperature (T=25000K) and gravity (log $g = 8.5$), consistent with the characteristics determined  from optical observations. However,  a small, significant excess is measurable for the K band, similar to that detected for "dusty" isolated white dwarfs harbouring suspected planetary debris. The possible existence of such circumstellar material around Sirius-B has still to be confirmed by further observations.

These deep images allow us to search for small but yet undetected companions to Sirius.
Apart from Sirius-B, no other source is detected within the total 25\arcsec\, field. A comparison of the flux expected from the faintest known brown dwarfs at the distance of Sirius demonstrates that the above limiting magnitudes correspond to a star of spectral type later than T5 at 5\arcsec\, and T7 at 10\arcsec. Using theoretical spectra of brown dwarfs and planet-size objects,  we also show that  the end of the brown dwarf sequence is reached in the outer part of the image. The minimum detectable mass is around 10 M$_{Jup}$ inside the planetary limit, indicating that an extrasolar planet at a projected distance of $\sim$ 25 AU from Sirius would have been detected.}
   {}


   \maketitle
%

\section{Introduction}

   Although Sirius is the brightest star in the sky, it is by no means an easy target for modern astronomy. Its extreme brightness (m$_v$=-1.46)  in fact presents significant problems for both observations and precise photometric in the immediate surroundings of the star. \\
Sirius is known to be a binary system since the prediction of a companion by Bessel in 1844 and subsequent observation  by Alvan Clark in 1862  of Sirius-B which worked out to be the closest white dwarf \citep[see][]{wesemael82}.
The system was also proposed to be triple because a visual companion was reported consistently around 1930  \citep[see][]{baize31} and persistent periodic ($\sim$ 6yr) residuals were also noticed in the A-B binary orbit \citep{volet32,benest95}. The existence of an interacting third star in an eccentric orbit was also proposed to explain the apparent historical change in color of Sirius \citep{gry90,bonnet91}.\\
Over many years, Sirius-B was monitored extensively in the optical
\citep{gatewood78}, although the stellar field around Sirius was virtually unknown till recently. Due to the high diffuse background produced by the bright Sirius-A, long exposures such as those of the Palomar plates generate a large $\sim$ 1\degr\, overexposed spot at the star position.
The first catalogue of stars in a (2.5x4)\arcmin\, field around Sirius was provided in an effort to isolate possible companion candidates \citep{bonnet91,bonnet00}. It enabled the identification of an unrelated m$_g$$\simeq$12 background star that was in close ($\sim$ 7\arcsec) conjunction with Sirius during the years around 1930, due to its high proper motion. This conjunction most likely explains the spurious companion reported at that time.\\
Modern techniques for data from space and ground-based observatories have allowed considerable progress to be achieved in the search and study of faint companions around bright stars.
\citet{schroeder00} imaged Sirius-A at 1.02 $\mu$m, with the Hubble Space Telescope (HST) Planetary Camera and provided first constraints within 17\arcsec\, of the star. \citet{kuchner00} using the HST-NICMOS camera in coronographic mode covered the central 3.5\arcsec\, at a similar 1.10 $\mu$m wavelength in a search of  exozodiacal dust around Sirius-A.
The HST-STIS spectrograph was also used to measure accurate UBVRI magnitudes of Sirius-B from its visual spectrum \citep{barstow05}.\\
Since these observations, ground-based coronographs using adaptive optics in the near infrared have emerged as a powerful new tool in searching for faint companions to nearby stars. We present the first JHK infrared images of a 25\arcsec\, field around Sirius-A acquired using such a device. The high constrast images allow the precise determination of Sirius-B infrared colors and provide the strongest constraints in the region 3-10 arcseconds from Sirius-A of the existence of a small companion in the Sirius system, down to a planetary size.  
%
  \begin{figure*}
  \sidecaption
   \centering
   \includegraphics[width=\textwidth]{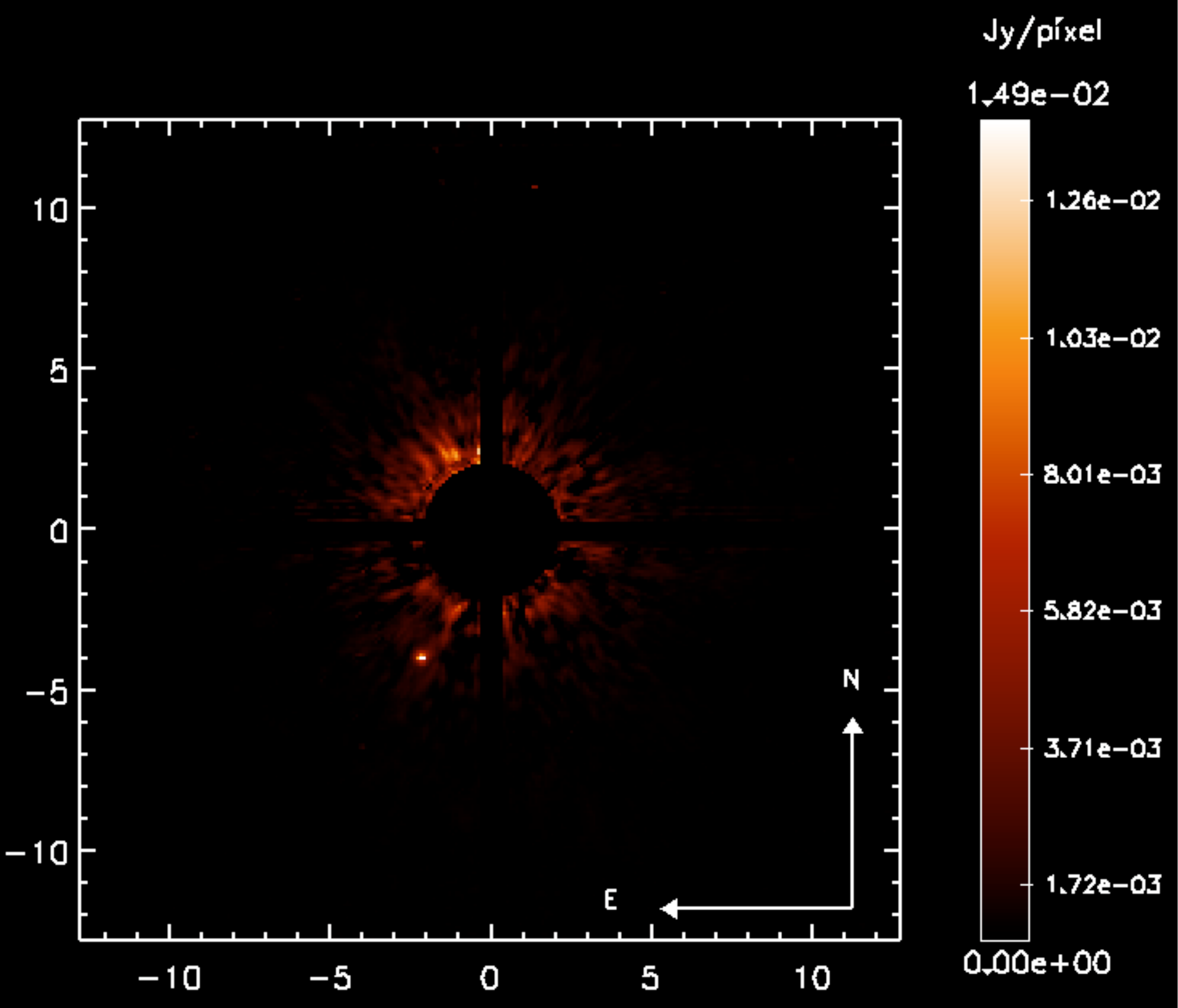}
      \caption{ADONIS Ks-band image of the white dwarf companion of Sirius. 
The pixel scale is 0.1\arcsec/pixel, the total field is
25.6x25.6\arcsec and the linear flux scale is given to the right. The Sirius star is centered under a 3.9\arcsec\, diameter coronographic mask and an additive numerical mask has been added to provide a clearer display. The white dwarf companion, Sirius-B, is clearly seen in the lower-left quadrant at a separation of 5.0\arcsec. }
       \label{figure1}
   \end{figure*}
%
\section{Observations}
Sirius was observed during two epochs from 2000 January, 14 to 16 and 2001
January, 13, using the SHARP II+ camera coupled with the
adaptive optics system ADONIS, and mounted onto the ESO 3.6m telescope at
La Silla, Chile \citep{rousset99}.  A pre-focal optics coronograph
\citep{beuzit97} was used to reject the direct starlight of Sirius-A
and increase the integration time in each elementary exposure.
Because of the high brightness of Sirius-A, we had to use a large
mask (diameter of 3.92\arcsec). The SHARP camera was used with a pixel scale of 100 mas  to increase the sensitivity to point-like faint companion and provide a total (25.6x25.6)\arcsec\, field of view.  
J (1.25 $\mu$m), H (1.64 $\mu$m), and Ks (2.15 $\mu$m) large band exposures were
obtained.  The seeing was quite stable during these observations at $\sim 1$\arcsec\, on average (on a
timescale of 1 night), but could reach a value
of 1.7\arcsec\, for H band data of January 2000.  We spent a total
observing time of 300 s in J band, 410 s  in H band and 800 s in the Ks band.  
The Point Spread Function (PSF) was monitored frequently by 
interlaced observations of two reference stars,
2 Cma (B1 II/III, m$_{\mbox{\tiny{V}}}$=1.98) and
$\gamma$ Cma (B8 II, m$_{\mbox{\tiny{V}}}$=4.097).
For selecting suitable reference stars and minimizing the PSF diffferences, three criteria were considered in order  of importance : the distance on the sky to the target,  the brightness match at the wavefront analysis wavelength (0.6 $\mu$m), and the spectral type match. The two reference stars are the most effective compromise and cover both the brightness and spectral type range.
These reference stars were
used later in the reduction process to subtract the wings of the
stellar PSF, and increase the sensitivity to a faint
companion. Typical FWHM of the PSF were 0.2\arcsec\, in J band and 
0.3\arcsec\, in H and Ks bands. Observations of empty fields were also performed to
estimate and remove the background flux, which can be significant in K
band observations. The reference stars HR 3018 and HD 19904 were used for
photometric calibration.

\section{Reduction procedure}
First, standard reduction techniques 
(including bias subtraction and flat-field correction) were applied to the data.  
For each filter, we obtained a
set of Sirius observations and corresponding PSFs. In spite of the use
of a coronograph mask, the image surface brightness was still dominated
by the stellar emission at any distance from the star.  To search
for faint companions, we had to subtract numerically the
starlight wings. The approximate subtraction of a scaled PSF to Sirius
images provides inaccurate results because of slight shifts (up to 1
pixel) in position on the array between the reference star and the object,
uncertainties in the fluxes (given by the literature), and a residual
background (ADONIS bench emission, different airmasses).  
We developed
a specific method to achieve an optimum subtraction.  

\subsection{Subtraction of the PSF}
For a pair of Sirius image ($\mbox{Obj}$) and corresponding PSF
($\mbox{Psf}$), we attempt to estimate automatically three parameters: a
shift ($\delta x$,$\delta y$) between the two images, a scaling factor
$R$, and a residual background $Bg$. These parameters are estimated by
minimizing the following error functional :
\begin{eqnarray}
 J & = & J_{\chi^2} \; + \; \alpha_{neg}J_{neg}\; + \;\alpha_{bal}J_{bal}\\ 
\mbox{where} & & \nonumber \\ 
 J_{\chi^2} & = & \sum_{\mbox{\tiny{subframe}}} \arrowvert \mbox{Obj} - S(\mbox{Psf}/R,\delta x, \delta y)- Bg \arrowvert\\
\mbox{where} & & \nonumber  \\ 
 J_{neg} & = & \sum_{\mbox{\tiny{negative pixels}}} \arrowvert \mbox{Obj} - S(\mbox{Psf}/R,\delta x, \delta y)- Bg \arrowvert \\ 
 \mbox{and} & & \nonumber \\ 
J_{bal} &=& \sqrt{\sum_i\sum_i(\mbox{med}(\mbox{quad}_i)-\mbox{med}(\mbox{quad}_j))^2}\\
\end{eqnarray} 
where [quad$_i$]  designates one of the 4 square quadrants around the star, 
the [S] function represents the image shift, and [med] function is the
median estimator. The sum of the $\chi^2$ term is performed over a
subframe located in a region of the images close to the star
(typically between circles of 25 and 50 pixels in radii), from which
unreliable pixel values are excluded (bad pixels, pixels belonging to areas
contaminated by diffracted light from the coronograph
support/telescope spider). The $J_{\chi^2}$ expresses the fidelity of
the shifted/rescaled PSF to the object image, the $J_{neg}$ prevents
overestimation of the ratio ($R$) parameter that would produce a
large, centered zone of negative pixels in the PSF subtracted result,
while, $J_{bal}$ prevents non uniformities in the four quadrants,
which are particularly high when the shift parameters are 
estimated incorrectly. We note that the central part of the images is not saturated and that bad pixels and the centermost region (where
the coronographic mask is located) are excluded from the computation
of the median value.  $\alpha_{neg}$ and $\alpha_{bal}$ are two weight
parameters with optimal values determined  experimentally as 1.0 and 2.0 respectively,
 so that the $\chi^2$ minimum value is of the order of the number of pixels.\\

The functional minimum is found using a zeroth-order minimization algorithm
called ``simplex'' \citep{press93}.  The method was first verified using 
simulated data whose input parameters (shifts, scaling factor) were
recovered with an accuracy of higher than 5 \%.  \\ 
To evaluate
the errors in the subtraction process and test the stability of the PSF,
the same subtraction process was applied using the two calibration
stars. Figure~\ref{figure1} shows the resulting subtracted image in the
Ks filter. Residuals level can be seen from random structures around
the coronographic mask. A point-like object can be seen easily in this
subtracted high contrast image. 
This is the first direct image of the white dwarf
companion of Sirius (pinpointed by the arrow) in this energy range.
The image has a sharper contrast than a comparable one produced 
using the WFPC2 camera on the HST telescope \citep{barstow05}, although 
the A-B magnitude difference does not differ significantly between the optical and infrared.
%
\begin{table}
\caption[ ]{JHK Photometry of Sirius-B}
     \label{table1}
\begin{flushleft}
\begin{tabular}{lrrrr}
\hline
\hline
\multicolumn{1}{c}{Band} & \multicolumn{1}{c}{Jan. 2000} & 
\multicolumn{1}{c}{Jan. 2001} & \multicolumn{1}{c}{Mean}
& \multicolumn{1}{c}{Abs. Mag.(*)} \\
\hline
	\noalign{\smallskip}
  	J (1.215$\mu$m) 		&  $9.11^{+0.08}_{-0.06}$ & $9.17^{+0.16}_{-0.13}$ 
		& $9.14^{+0.12}_{-0.09}$ & $12.03^{+0.12}_{-0.09}$ \\ 
   	\noalign{\smallskip}
  	H (1.654$\mu$m) 		&  $9.34^{+0.36}_{-0.24}$ & $9.02^{+0.14}_{-0.09}$ 
	& $9.17^{+0.23}_{-0.16}$ & $12.06^{+0.23}_{-0.16}$  \\ 
  	\noalign{\smallskip}
	$\rm K_{ \rm s}$ (2.157$\mu$m)  &  $9.12^{+0.29}_{-0.23}$ & $8.91^{+0.09}_{-0.07}$ 
	& $9.01^{+0.18}_{-0.14}$ & $11.90^{+0.18}_{-0.14}$ \\ 
 \noalign{\smallskip}
\hline
\end{tabular}
\end{flushleft}
(*) using the Hipparcos parallax $\pi$=0.3791 \arcsec
\end{table}
%
%
  \begin{figure}
   \centering
   \includegraphics[width=8.9cm,angle=-0]{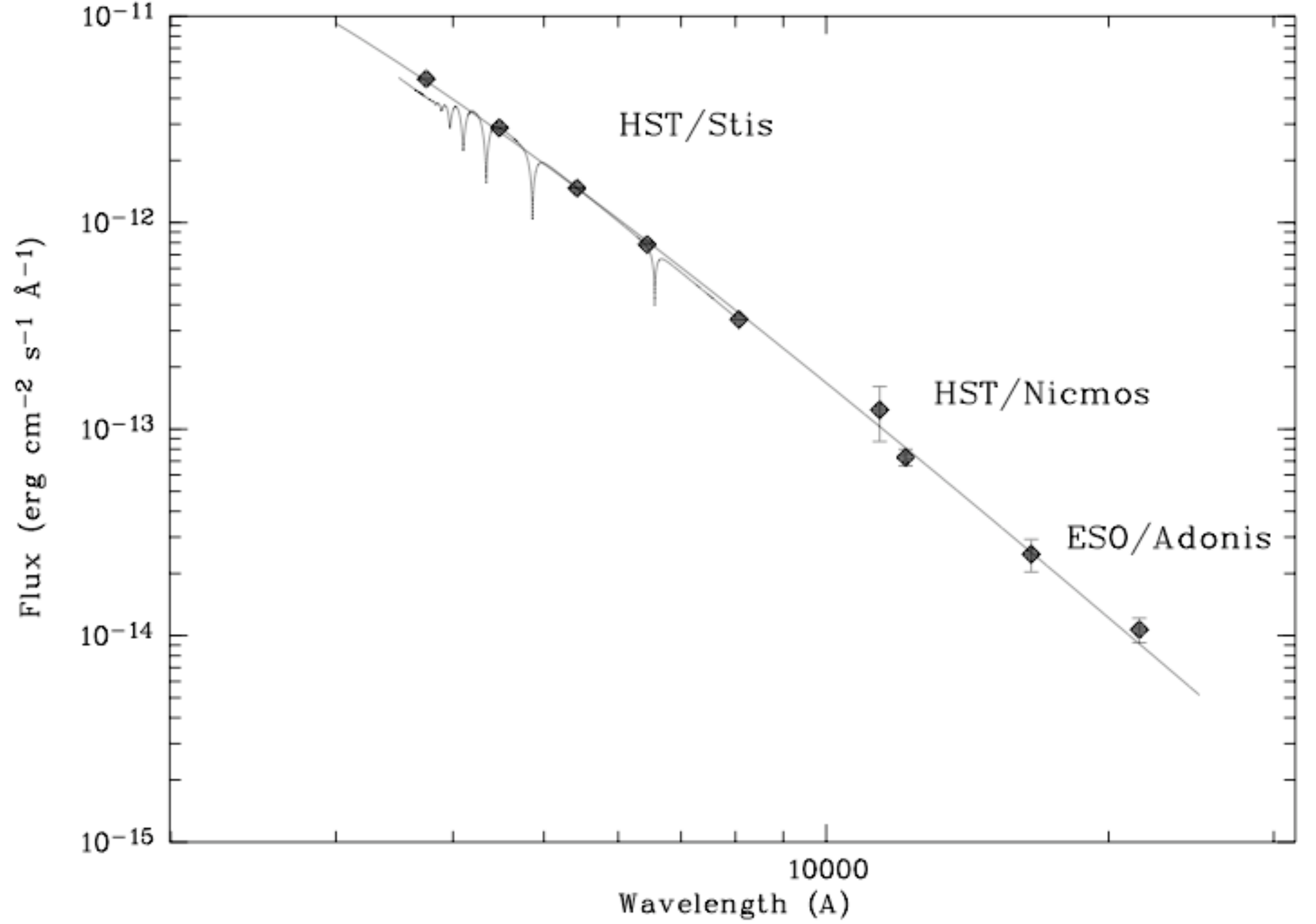}
      \caption{Sirius-B energy distribution from optical to infrared.   
Monochromatic fluxes (black dots) are from UBVRI (HST-STIS), 1.104 $\mu$m (HST-NICMOS) and JHKs magnitudes (this work).
Also shown in the optical is the best Koester theoretical model (see text). The full line corresponds to a pure blackbody at a T=25\,190K temperature. See text for a more detailed comparison to theoretical models in the infrared.              }
       \label{figure2}
   \end{figure}
%
\section{Infrared photometry of Sirius-B}
In spite of the high image quality, the photometric measurements for the companion 
observed in Fig.~\ref{figure1} is difficult. To obtain accurate photometric measurements, 
special care must be taken to remove systematic effects due to light contamination.
 The high level of surrounding
residuals prevents the use of standard photometric estimation methods
such as encircled energy summation.  After assessing many methods using
simulated data, the most accurate one proved to be a method based on PSF
fitting of the source. It is unaffected by the surrounding residual
structures produced by the PSF subtraction, and depends weakly on
uncertainties when estimating the background in the image. The
photometric measurements of the WD are given in
Table~\ref{table1} for the two epochs. The uncertainties given in this table were
estimated by combining the actual surrounding residual uncertainties
and the results of simulations based on Monte Carlo trial processes.
Fluxes were converted into magnitudes using the filter responses and zero-points
defined by \citet{tokunaga00} in the MKO system, which do not 
differ significantly from the calibration given by \citet{cohen03b}. The last column of  
Table~\ref{table1} provides the average absolute magnitudes computed using the parallax distance 
determined by the Hipparcos satellite \citep{esa97}. These are the first accurate JHK magnitudes for Sirius-B, which can be compared to theoretical expectations. 

The optical photometry as well as the temperature and gravity of the white dwarf (logT= 25\,190K, logg=8.556) were determined accurately using HST Balmer lines spectroscopy (Barstow et al. 2005). 
Figure ~\ref{figure2} shows the measured energy distribution of Sirius-B from optical to infrared,
compared to the flux distribution from a pure blackbody at T=25\,190K, scaled to the optical flux at  5500\AA. 
Also shown in the optical range is the synthetic WD spectrum interpolated in temperature and gravity from a grid of WD models with LTE atmospheres (\citet{finley97}, Koester (2000) private communication).
UBVRI fluxes were converted to the magnitude system of \citet{cohen03a}. The flux measured at 1.104 $\mu$m from HST-NICMOS observations (Kuchner \& Brown 2000) is also shown. The JHKs infrared fluxes are already in remarkable agreement with this simple black body extrapolation.

In the infrared range, comparison to more detailed synthetic colors of DA white dwarfs was performed by interpolating the grid of synthetic photometry provided by \citet{holberg06} on the basis of LTE model atmospheres.  In these new models, JHKs magnitudes are computed in the filters and magnitude scale of \citet{cohen03b}, which are equivalent to our measurements. 
Considering the remaining uncertainities in the WD mass and radius (Barstow et al. 2005), the model magnitudes for a pure H atmosphere white dwarf at the Sirius-B temperature and gravity were scaled to the V absolute magnitude (M$_v$=11.422). The theoretical predictions for the Sirius white dwarf magnitudes are then
M$_J$=12.033, M$_H$=12.120, and M$_{Ks}$=12.213 which yield "observed-theoretical" magnitude differences of 
$0.001^{+0.12}_{-0.09}$, $-0.058^{+0.23}_{-0.16}$,  and $-0.309^{+0.18}_{-0.14}$, respectively in the JHKs bands. 
The error bars were computed here not from an "assumed" normal distribution but from the true distribution of residuals amplitudes, derived from the image statistics in the related region. They
 correspond to an exclusion of a false excess detection at a 99,64\% confidence level.  
Whereas the J and H magnitudes reproduces accurately the predicted values, the K magnitude
has a small but significant excess of $\sim 0.3$ magnitude (see Figure ~\ref{figure2}). 
Interestingly, a similarly small K excess was also recently measured for selected cool (T  $\leq$ 12 000K), isolated white dwarfs for which an overall excess of flux at wavelength longer than 2µm with characteristics consistent with circumstellar dust or debris  is found  \citep{vonhippel07}.
In a Spitzer mid-infrared survey of 124 white dwarfs, four "dusty" white dwarfs were found
with dusty environment that may represent the remains of
planetary systems \citep{reach05} and 
a metal-rich gas disk was discovered around a hotter (T=22
500K) WD, possibly associated with planetary debris material 
\citep{gaensicke08}. It is therefore possible that the small departure observed in the K band indicates a similar circumstellar material around Sirius-B.
%
%
  \begin{figure}
   \centering
   \includegraphics[width=8.5cm,angle=-0]{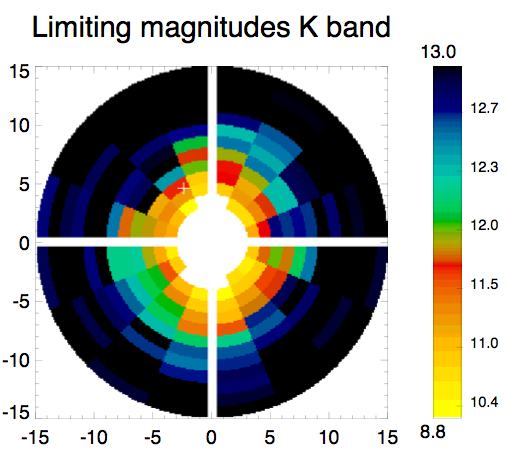}
      \caption{Azimutal variation of the limiting magnitudes in the $\rm K_{ \rm s}$ band image for a point source in a 15\arcsec field around Sirius (scale in arcsec). The vertical right color scale gives the range of limiting apparent magnitude for the different sectors. }
       \label{figure3}
   \end{figure}
%
\section{Looking for a third star}
The high contrast images obtained here in the near infrared are useful  for constraining the existence of a possible small mass companion in the Sirius system. 
No point-source other than Sirius-B can be detected in the field, down 
to a certain limiting sensitivity that depends on the filter used. 
We estimate our limiting magnitudes in each filter (J,H,Ks) by simulating
a point source hidden in the residuals of the PSF subtracted image. 
The minimum detectable magnitude in the different regions of the image was computed 
by evaluating the standard deviation and cumulative probability  of the residuals in each sector.   
The detection limit was set from the true probability to correspond to  a  significance $P \geq 0.9$,
which in our case corresponds to a 10 $ \sigma$ limit for the peak intensity detection and a
500 $ \sigma$ limit for a PSF integrated intensity. 

\begin{figure}[htb]
\centering
\includegraphics[width=8.5cm,height=6.5cm,clip=,angle=0]{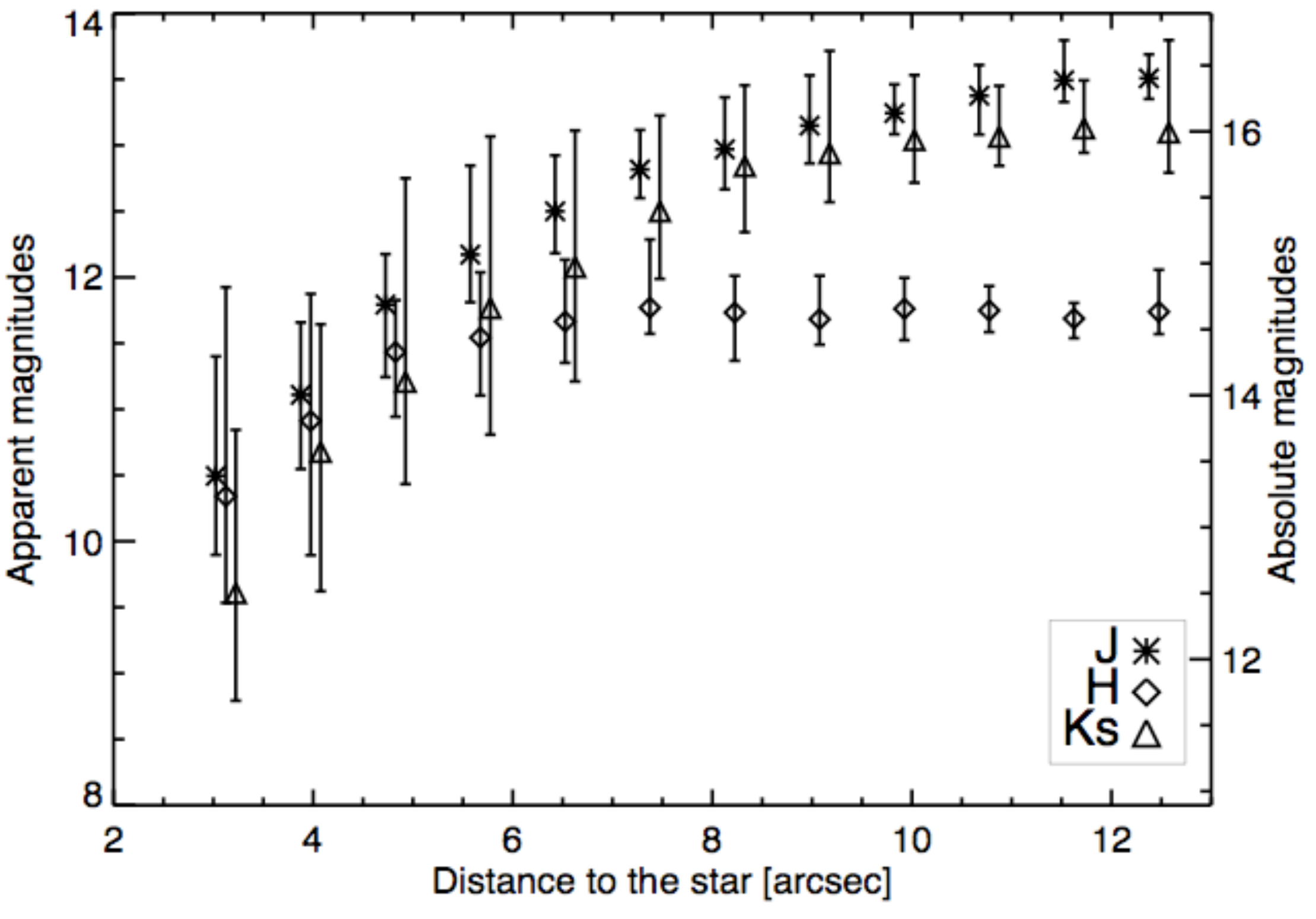}
\caption{Plot of the limiting sensitivities in magnitudes as a function of the angular separation to the star 
in the three near-infrared bands (crosses : J, open lozenges: H, open triangles: $\rm K_{ \rm s}$). Points are slightly shifted horizontally for clarity. Note that the vertical lines are not statistical error bars but mark  the minimum/maximum range of azimuthal variations at a given radius.The right scale gives the absolute magnitude at the Sirius distance.}
\label{figure4}
\end{figure}

\begin{figure}
   \centering
   \includegraphics[width=8.6cm,angle=0]{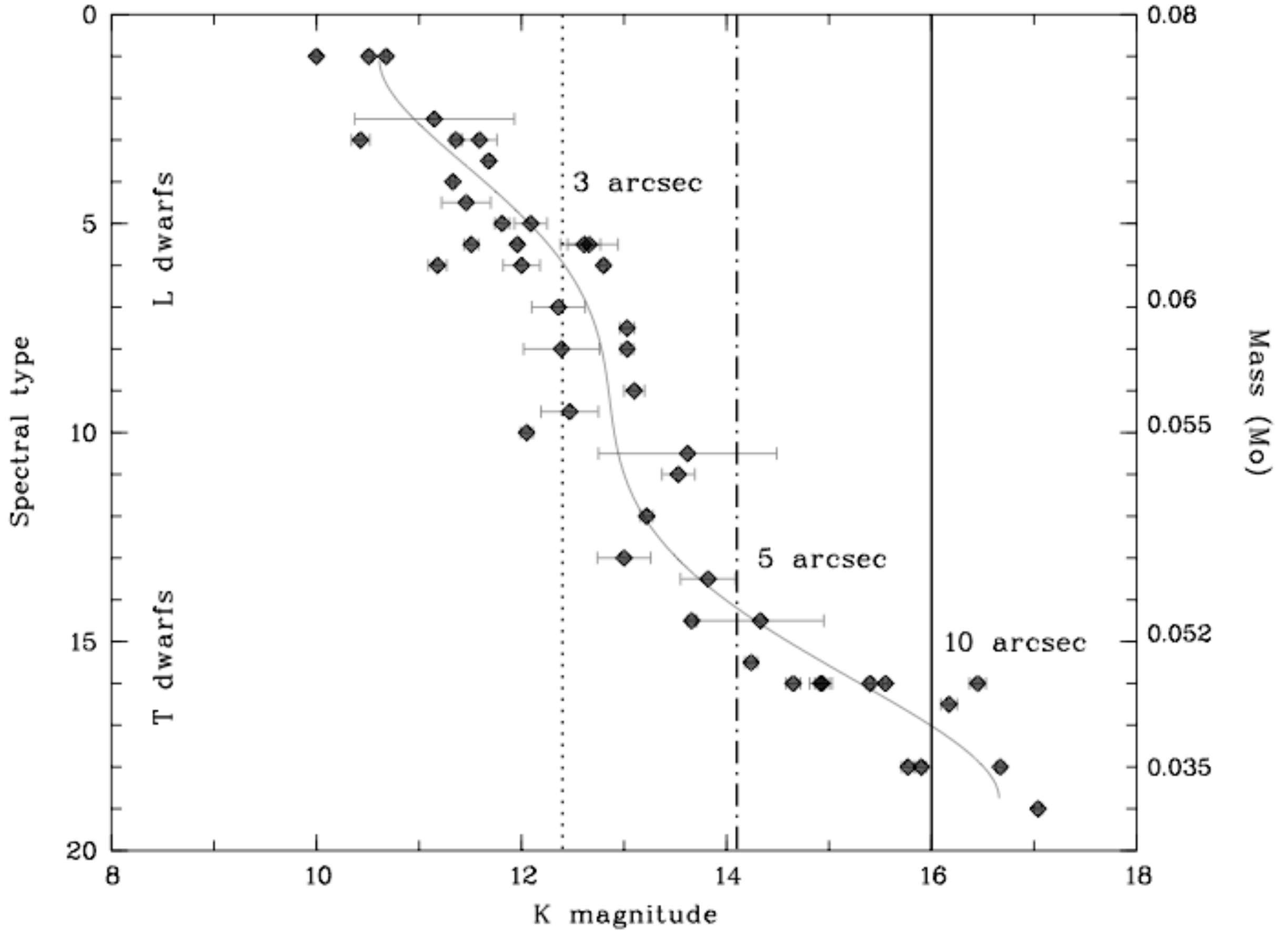}
      \caption{Limiting K absolute magnitude at different angular separations from Sirius-A 
(10 \arcsec : thick line - 5 \arcsec : dash-dotted line - 3 \arcsec : dotted line)
compared to the faintest observed L and T dwarfs (black dots). Spectral types (left scale)
are coded by 01=L01, 10=T0, 19=T9 and representative masses are 
indicated on the right-hand side axis.
The thin line is the best M$_K$-spectral type fit from \citet{knapp04}. 
}
 \label{figure5}
 \end{figure}

Figure ~\ref{figure3} shows the results for the $\rm K_{ \rm s}$ image. 
Significant azimuthal variations of up to $\sim$1 mag. are present in the image with variations decreasing
toward the outer part of the image as the level of PSF subtraction residuals decreases with
the distance to Sirius.  Typical  limiting sensitivities as a function of distance to Sirius were derived by azimuthally averaging the residual levels of the PSF subtracted image and are shown in Figure ~\ref{figure4} for the different filters.
%
%
%
The most robust constraints are obtained for the J and $\rm K_{ \rm s}$ filters. 
For the filter $\rm K_{ \rm s}$, the upper limits range in absolute magnitude 
from M$_{Ks}$$\sim 12.4$ at 3\arcsec\, 
of Sirius-A to M$_{Ks}$$\sim 16.0$ at 10\arcsec.
At these levels, M dwarfs are already excluded. Using an extrapolation of
the empirical mass-luminosity  relations estimated by \citet{delfosse00},  
M$_{Ks} \geq 10$ as in Figure ~\ref{figure4} corresponds to a mass M$\leq 0.08$ M\sun.
These magnitudes are only comparable with those observed for the faintest dwarfs known.
In the last ten years, hundreds of L and T dwarfs have been discovered.
We used the catalog of 71 L and T dwarfs of \citet{knapp04}, in which 45 have known distance and therefore absolute K magnitudes (see their table 8). 
Figure ~\ref{figure5} shows this selected observed sample with our magnitude upper limits at the different distances from Sirius-A. Using the polynomial fit of \citet[ table 12]{knapp04}, the upper limits at 5\arcsec\, and 10\arcsec\, correspond to spectral types later than T4.8 and T7.0, respectively.
There is no simple mass-luminosity relation for L and T dwarfs since many parameters (gravity, age, metallicity,..) are involved  \citep[see][]{burrows06} but a approximate estimation can be obtained from
distributions computed from various models \citep{burgasser04}. The upper limit at 10\arcsec\, corresponds to a companion mass in the range M=(0.042-0.052) M\sun\, or (44-54) M$_{Jup}$, down to the brown dwarf range and close to planetary limits. %
\begin{table}
\caption[ ]{Limiting Ks magnitudes and masses around Sirius}
     \label{table2}
\begin{flushleft}
\begin{tabular}{lrrr}
\hline
\hline
\multicolumn{1}{l}{Separation} & \multicolumn{1}{r}{3\arcsec} & 
\multicolumn{1}{r}{5\arcsec} & \multicolumn{1}{r}{10\arcsec} \\ 	
\multicolumn{1}{l}{Distance (AU)} & \multicolumn{1}{r}{7.9} & 
\multicolumn{1}{r}{13.2} & \multicolumn{1}{r}{26.4} \\
\hline
	\noalign{\smallskip}
	m$_{Ks}$(obs)		&  9.5 & 11.2 & 13.1  \\ 
  	M$_{Ks}$(obs)		&  12.4 & 14.1 & 16.0  \\ 	
   	\noalign{\smallskip}
	\multicolumn{4}{l}{ \it{Models*}}\\	
  	M$_{Ks}$(models)	&  13.1 & 14.3 & 16.6  \\
	Mass (M$_{Jup})$ &  30    &  20    & 10    \\
 \noalign{\smallskip}
\hline
\end{tabular}
\end{flushleft}
(*) Burrows et al. (2006) models
\end{table}
%
Independent estimations can also be obtained using theoretical L-T dwarfs  models. 
We used published spectral models for brown dwarf  \citep{burrows06} and planetary  \citep{burrows03} masses to compute the expected infrared magnitudes.  We selected the models with solar abundances and an age of 250 Myr appropriate for Sirius \citep{liebert03} and derived the temperature and gravity corresponding to a given mass, using the Burrows Brown Dwarf and Extra-Solar Giant Planet Calculator. The corresponding theoretical spectra were then convolved with the filter response to compute the magnitudes. 
Table~\ref{table2}  gives the closest masses compatible with the observed upper limit at different projected distances from Sirius. From our K-image, a third star with a mass $\sim$ 30 M$_{Jup}$ could be detected close to Sirius ($\sim$ 8 AU), while a miminum mass of $\sim$ 10 M$_{Jup}$ is reached at a distance of $\sim$ 26 AU. This final limit indicates that an extrasolar planet around Sirius at a distance comparable to the Sun-Neptune distance would have been detected. These are the best constraints in a (4-25\arcsec) region around Sirius-A.
From the published HST-NICMOS upper limit  at a shorter (1.1 $\mu$m) wavelength \citep{kuchner00}, we used the same above method to derive a limit  of $\sim$ 45 M$_{Jup}$ and $\sim$ 15 M$_{Jup}$, closer to Sirius, at a separation of 2\arcsec\, and 3\arcsec\,, respectively.
 With the negative optical search in the wider (2.5x4\arcmin) field \citep{bonnet00}, this weakens considerably the possibility of a third star in the Sirius system.

The high resolution achieved by adaptive optics also allows the search for a suspected faint star in a close orbit around Sirius-B. From an analysis of the orbit residuals, a $\sim$ 6yr periodicity is present and a general three-body model indicates that possible stable orbits exist with a restricted range of masses M$\leq 0.038$ M\sun\, (40 M$_{Jup}$) and semi-major axis a$_0$=(1-2.5) AU \citep{benest95}. The reconstructed PSF of FWHM = 0.31 +/-0.05\arcsec\, in the Ks band is equivalent to a 0.8 AU separation from Sirius-B at the system distance therefore the companion could be resolved in our image. At the position of Sirius-B (5\arcsec), the upper limit in our K image  is M$_{Ks}=14.1$. This corresponds to a theoretical mass of $\sim$ 20 M$_{Jup}$ (Table~\ref{table2}), lower than the predicted mass. Our limit therefore also excludes a suspected faint component to Sirius-B unless the orientation is very unfavourable.
%
%
%


\section{Conclusions}
The infrared image of the Sirius field presented here is the first high constrast image in the JHK wavelength range. Despite the coronagraphic device associated with adaptive optics, the  light contamination of the brightest Sirius dominates the field. A precise subtraction of the diffuse background and a careful calibration enable us however to derive very accurate constraints on the different objects in the field. 
The infrared absolute magnitudes of Sirius-B are determined to be  M$_J$=$12.03^{+0.12}_{-0.09}$, M$_H$=$12.06^{+0.23}_{-0.16}$, and M$_{Ks}$=$11.90^{+0.18}_{-0.14}$. The JH values are  in excellent agreement with the white dwarf theoretical model of a DA white dwarf at a temperature and gravity determined accurately from the HST image. A small departure is visible in the K band, which may indicate a possible  circumstellar material around Sirius-B, similar to that observed around some selected "dusty" white dwarfs. This has yet to be confirmed by observations at longer ($2-15\mu$m) wavelengths, where most of the "dust" emission is expected.
The high quality image also allows a deep search for possible low-mass objects in the field. Although the residual background after subtraction shows significant azimuthal variations,  mean limiting magnitudes in the field reach the planetary limit for an object located at the Sirius distance.
The deep field obtained here around Sirius provides a limit of (30-10) M$_{Jup}$ in the (8-26) AU region and complementary HST-NICMOS observations yield a similar limit down to 5 AU.  Since the most central part of the image ($\leq$5 AU) has still not been covered, this does not fully eliminate the possibility of a third member in the sytem but the probability of a triple system  is now low.

\begin{acknowledgements}
      We thank the 3.6m ESO team for their helpful support during the observations and F. Marchis and O. Marco for useful advices. We also thank an anonymous referee for very useful comments and insightful remarks to improve the manuscript.
\end{acknowledgements}

%

\begin{thebibliography}{}

   \bibitem[Baize(1931)]{baize31} Baize, P., 1931, L'Astronomie 45, 383-397
    \bibitem[Barstow et al.(2005)]{barstow05} Barstow, M. A.; Bond, H. E.; Holberg, J. B. et al., 2005, MNRAS 362, 1134
  \bibitem[Benest \& Duvent (1995)]{benest95} Benest, D..; Duvent, J.-L., 1995, A\&A, 299, 621
  \bibitem[Beuzit et al.(1997)]{beuzit97} Beuzit, J.-L.; Mouillet, D.; Lagrange, A.-M.; Paufique, J., 1997, A\&AS, 125, 175
  \bibitem[Bonnet-Bidaud \& Gry(1991)]{bonnet91} Bonnet-Bidaud, J.M., Gry, C., 1991, A\&A 252, 193
  \bibitem[Bonnet-Bidaud, Colas \& Lecacheux(2000)]{bonnet00} Bonnet-Bidaud, J.-M.; Colas, A.; Lecacheux, J., 2000, A\&A, 360, 991
   \bibitem[Burrows et al.(2003)]{burrows03} Burrows, A.; Sudarsky, D.; Ludine, J., 2003, ApJ, 596, 587
  \bibitem[Burrows et al.(2006)]{burrows06} Burrows, A.; Sudarsky, D.; Hubeny, I., 2006, ApJ, 640, 1063
  \bibitem[Burgasser (2004)]{burgasser04} Burgasser, A., 2004, ApJS, 155, 191
  \bibitem[Cohen et al.(2003a)]{cohen03a} Cohen, M.; Megeath, S. T.; Hammersley P.; Martin-Luis F.; Stauffer J., 2003, AJ, 125, 2645 
  \bibitem[Cohen et al.(2003b)]{cohen03b} Cohen, M.; Wheaton, W. A.; Megeath, S. T., 2003, AJ, 126, 1096
  \bibitem[Delfosse et al.(2000)]{delfosse00} Delfosse, X.; Forweille, T.; Segrensan, D. et al., 2000, A\&A, 364, 217
  \bibitem[ESA(1997)]{esa97} ESA 1997, The Hipparchos and Tycho catalogues, ESA SP-1200, Noordwijk
 \bibitem[Finley et al.(1997)]{finley97} Finley D.S., Koester D., Basri G., 1997, ApJ 488, 375
  \bibitem[Gatewood \& Gatewood(1978)]{gatewood78} Gatewood, G., Gatewood, C., 1978, ApJ 225, 191
  \bibitem[Gaensicke et al.(2008)]{gaensicke08}Gaensicke B.T., Marsh T.R., Southworth J., Rebassa-Mansergas A., 2008, Science 314, 1908
  \bibitem[Gry \& Bonnet-Bidaud(1990)]{gry90} Gry, C., Bonnet-Bidaud, J.M., 1990, Nat. 347, 625
   \bibitem[Holberg \& Bergeron(2006)]{holberg06} Holdberg, J. B.; Bergeron, P., 2006, AJ 132, 1121
  \bibitem[Kuchner \& Brown(2000)]{kuchner00} Kuchner, M. J.; Brown, M. E, 2000, PASP 112, 827
  \bibitem[Knapp et al.(2004)]{knapp04} Knapp, G.R.; Leggett, S. K.; Fan, X. et al., 2004, ApJ, 127, 3553
  \bibitem[Liebert et al.(2003)]{liebert03} Liebert, J., Young P., Arnett D. et al., ApJ 630, 692
  \bibitem[Press(1993)]{press93} Press, W.H., 1993, in Numerical Recipes, ed.\ Camb. Univ. Press
  \bibitem[Reach et al.(2005)]{reach05} Reach W.T., Kuchner M., von Hippel T. et al., 2005, ApJ 635, L161
  \bibitem[Rousset \& Beuzit(1999)]{rousset99} Rousset, G.; Beuzit, J.-L., Adaptive optics in astronomy, ed F. Roddier. Cambridge, 1999
  \bibitem[Schoeder et al.(2000)]{schroeder00} Schroeder, D. J.; Golimowski, D. A.; Brukardt, R. A. et al., 2000, AJ 119, 906
  \bibitem[Tokunaga(2000)]{tokunaga00}  Tokunaga, A.T., in Allen's Astrophysical Quantities, 4th edition, ed. A.N. Cox, Springer-Verlag, NY, p. 143 (2000)
  \bibitem[Volet(1932)]{volet32} Volet, Ch, 1932, Bull. Astron. Paris  8, 51-64
    \bibitem[von Hippel et al.(2007)]{vonhippel07} von Hippel T., Kuchner M., Kilic M., Mullally F., and Reach W.T., 2007, ApJ 662, 554
  \bibitem[Wesemael \& Fontaine(1982)]{wesemael82} Wesemael, F., Fontaine, G., 1982, J. R. Astron. Soc. Can. 76, 35

\end{thebibliography}

\end{document}